\documentclass[conference]{IEEEtran}
\IEEEoverridecommandlockouts
\usepackage{cite}
\usepackage{amsmath,amssymb,amsfonts}
\usepackage{algorithmic}
\usepackage{graphicx}
\usepackage{textcomp}
\usepackage{comment}
\usepackage{xcolor}
\usepackage{multirow}
\usepackage{subcaption}
\def\BibTeX{{\rm B\kern-.05em{\sc i\kern-.025em b}\kern-.08em
    T\kern-.1667em\lower.7ex\hbox{E}\kern-.125emX}}

\usepackage{tikz}
\usepackage{textcomp}
\usepackage[doipre={DOI:~}]{uri}
\usepackage{lipsum}
\newcommand\copyrighttext{%
  \footnotesize \textcopyright 2022 IEEE. Personal use of this material is permitted.  Permission from IEEE must be obtained for all other uses, in any current or future media, including reprinting/republishing this material for advertising or promotional purposes, creating new collective works, for resale or redistribution to servers or lists, or reuse of any copyrighted component of this work in other works. 

  Accepted as a conference paper at the IEEE 2022 PRIME Conference.}
\newcommand{\copyrightnotice}{%
\begin{tikzpicture}[remember picture,overlay]
\node[anchor=south,yshift=10pt] at (current page.south) {\fbox{\parbox{\dimexpr\textwidth-\fboxsep-\fboxrule\relax}{\copyrighttext}}};
\end{tikzpicture}%
}

\begin{document}

\title{Two-stage Human Activity Recognition on Microcontrollers
with Decision Trees and CNNs
}

\author{
\IEEEauthorblockN{Francesco Daghero}
\IEEEauthorblockA{
\textit{DAUIN, Politecnico di Torino}\\
Turin, Italy \\
francesco.daghero@polito.it}
\and
\IEEEauthorblockN{Daniele Jahier Pagliari}
\IEEEauthorblockA{
\textit{DAUIN, Politecnico di Torino}\\
Turin, Italy \\
daniele.jahier@polito.it}
\and
\IEEEauthorblockN{Massimo Poncino}
\IEEEauthorblockA{
\textit{DAUIN, Politecnico di Torino}\\
Turin, Italy \\
massimo.poncino@polito.it}
}
\IEEEoverridecommandlockouts
\maketitle
\copyrightnotice%

\begin{abstract}
Human Activity Recognition (HAR) has become an increasingly popular task for embedded devices such as smartwatches.
Most HAR systems for ultra-low power devices are based on classic Machine Learning (ML) models, whereas Deep Learning (DL), although reaching state-of-the-art accuracy, is less popular due to its high energy consumption, which poses a significant challenge for battery-operated and resource-constrained devices. 
In this work, we bridge the gap between on-device HAR and DL thanks to a hierarchical architecture composed of a decision tree (DT) and a one dimensional Convolutional Neural Network (1D CNN).
The two classifiers operate in a cascaded fashion on two different sub-tasks: the DT classifies only the easiest activities, while the CNN deals with more complex ones.
With experiments on a state-of-the-art dataset and targeting a single-core RISC-V MCU,
we show that this approach allows to save up to 67.7\% energy w.r.t. a ''stand-alone'' DL architecture at iso-accuracy. 
Additionally, the two-stage system either introduces a negligible memory overhead (up to 200 B) or on the contrary, reduces the total memory occupation.

\end{abstract}


\section{Introduction and Related Works}\label{sec:intro}
Human Activity Recognition (HAR) based on Inertial Measurements Units (IMUs) has become an increasingly popular feature for smart devices such as fitness trackers and smartwatches.
A HAR task consists of determining the activity performed by an user in a specific time-window of the input signal. 
This is generally performed through Machine Learning (ML)~\cite{daghero2021ultracompact,comparison_hapt_rf} and more recently Deep Learning (DL)~\cite{daghero2021ultracompact,comparison_hapt_dnn}, with the latter generally achieving state-of-the-art accuracy.

As for many other tasks in the Internet of Things (IoT) world, HAR benefits from an edge-centric approach, that is, from performing the full ML inference on the same device collecting the sensors readings.
In this way, there is no need to transmit the data over possibly unstable or unreachable networks, thus making the inference latency more predictable. 
Additionally, possibly sensitive information don't have to be transmitted, increasing the application security. 
Finally, avoiding the data transmission brings an additional advantage in terms of energy efficiency: wireless connectivity is in fact power-hungry, generally consuming much more than performing the computations locally.
This is particularly beneficial since edge devices are often battery-powered.

However, directly deploying a HAR ML model tailored for the cloud on an embedded device is often difficult. 
Those devices are in fact typically based on ultra-low-power Microcontrollers (MCUs) with tight memory constraints and relatively low clock frequencies.
For this reason, most HAR implementations on embedded devices are based on simple classifiers such as Decision Trees (DTs).
These shallow ML models require only a few compare-and-branch operations for inference, and resulting in low latency and energy consumption. Additionally, their memory requirements are also orders of magnitude less than other types of classifiers. However, they rarely reach very high accuracies on complex tasks.

On the other hand, DL approaches have shown promising results for HAR~\cite{daghero2021ultracompact, comparison_hapt_dnn, comparison_hapt_rf}, achieving state-of-the-art accuracies (significantly higher than tree-based models) on several datasets. However, most DL models proposed in other works are impossible to deploy on MCUs, requiring $>1$ million parameters and
floating point operations for inference~\cite{comparison_hapt_dnn}. This, in turn, translates into unacceptable latency and energy consumption for IoT devices.

To bridge this gap, several works have introduced optimizations aimed at reducing the complexity of DL models and making them compatible with edge devices.
Among the most popular there are quantization and pruning~\cite{Jacob2018,molchanov2016pruning}, which modify the models statically before deployment, either at training time or post-training, respectively reducing the complexity of the network through the use of low-precision arithmetic or removing redundant parameters.

Orthogonal to such approaches, \textit{dynamic} (or \textit{adaptive}) optimization techniques leverage the varying complexity of inputs in order to save energy at runtime.
The key intuition is that performing an inference with an accurate yet energy demanding model is wasteful for easy inputs, while naively shrinking the classifier may lead to a deterioration in predictions' quality for hard inputs.
Additionally, easy inputs are often far more common than hard ones in real-word situations (e.g. for HAR the activity ''laying'' is far more frequent than ''walking upstairs''), thus making adaptive approaches even more effective in terms of energy efficiency.

Most existing dynamic inference systems are based on sequentially executing multiple models of increasing complexity on each input, all performing the \textit{same task}, and stopping the process as soon as a sufficient \textit{output confidence} is reached. The multiple models can be separate~\cite{park2015big} or share weights, using variable-width~\cite{tann2016runtime}, variable-precision~\cite{JahierPagliari2018a}, or early-exit~\cite{teerapittayanon2016branchynet,Daghero2021b} mechanisms.
Alternatively, so-called “hierarchical classifiers” \textit{split an inference task into different sub-tasks} based on the input. An example of this approach is~\cite{hierarchical2018}, which introduces a framework based on several cascaded classifiers of increasing complexity, each working on an increasingly more difficult sub-task, constructed so that easier and more common inputs stop earlier in the cascade, reducing the average energy consumption.

In this paper, we follow the latter approach, proposing a novel type of adaptive architecture tailored for HAR tasks. Specifically, we combine a DT and a 1-dimensional (1D) Convolutional Neural Network (CNN). 
The first is trained to select only among the easiest activities, and output a generic \textit{fallback} class for all other inputs.  Only when the fallback class is predicted by the DT, the CNN is activated, performing an inference pass on the same input.
We test our approach on a single-core RISC-V MCU and on a state-of-the-art HAR dataset~\cite{hapt}, showing that we are able to save up to 67.7\% energy at iso-accuracy, and with a negligible memory overhead, with respect to a static DL approach. More generally, our adaptive system is Pareto optimal when compared to multiple static CNN architectures, both in the energy versus accuracy and memory versus accuracy planes.

\section{Proposed Method}\label{sec:methods}
\begin{figure}[t]
    \centering
    \includegraphics[width=0.7\linewidth]{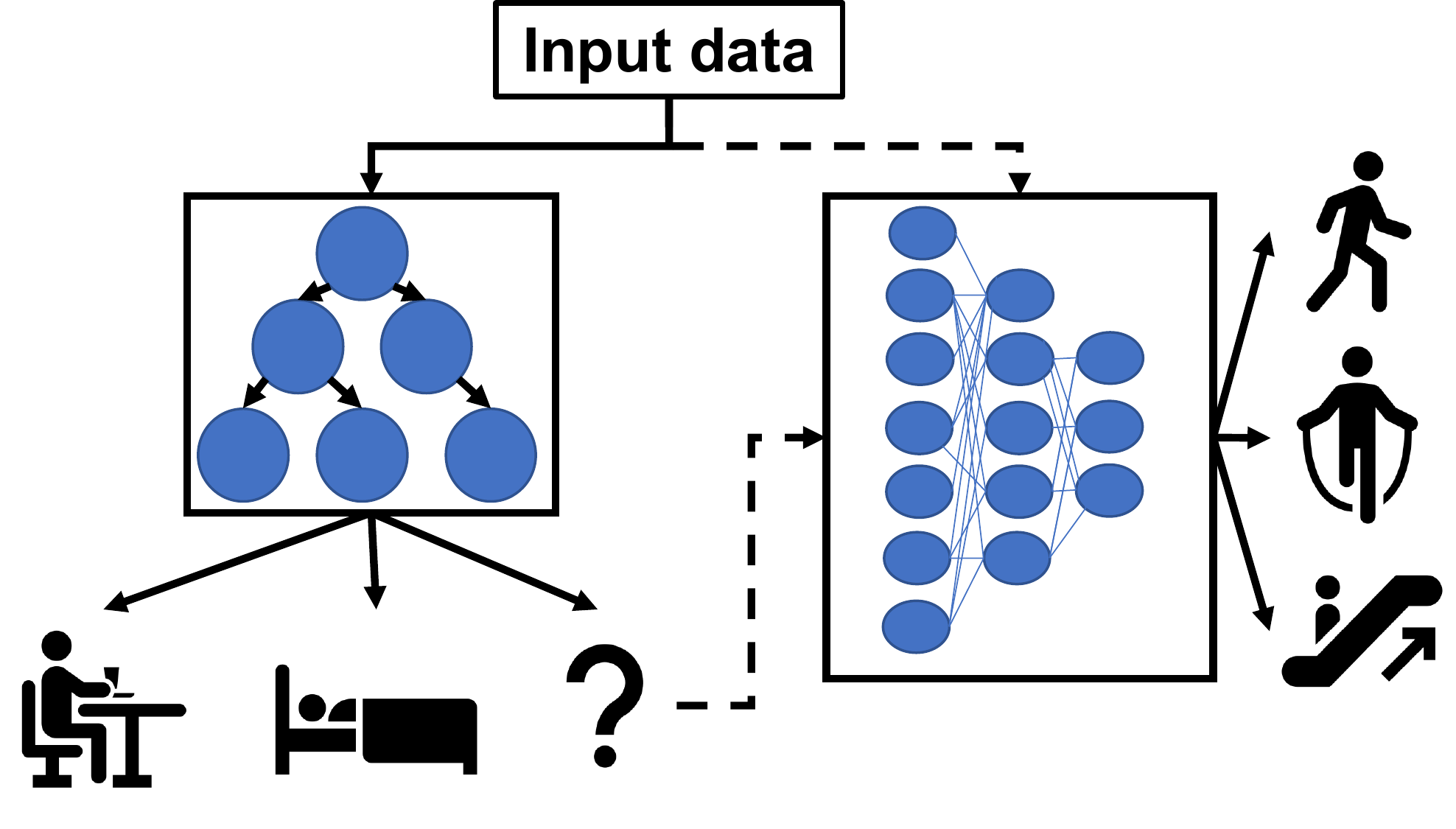}
    \caption{High-level overview of the two-stage hierarchical inference architecture. Easy classes are classified by the DT (left), while complex ones are left to the CNN (right).}
    \label{fig:architettura}
\end{figure}

\begin{figure*}[ht]
    \centering
    \includegraphics[width=0.9\linewidth]{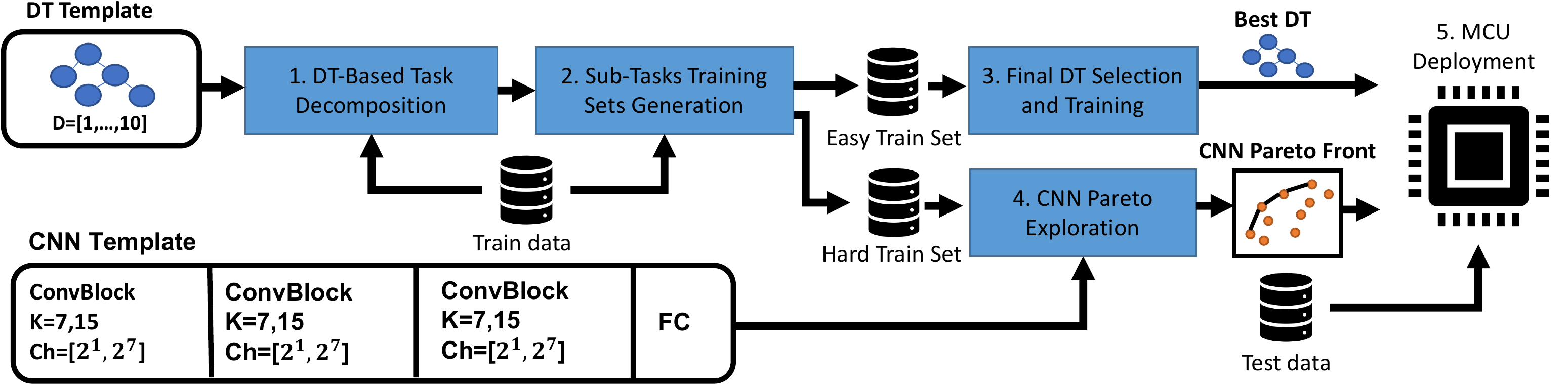}
    \caption{Schematic view of the proposed training and deployment flow.}
    \label{fig:workflow}
    \vspace{-3mm}
\end{figure*}

This work proposes the first two-stage hierarchical, hybrid ML/DL system for HAR, to our knowledge.

Fig.~\ref{fig:architettura} shows an high-level scheme of the solution, which mixes a DT with a 1D-CNN, each working on a different sub-task, rather than on the whole set of activities.
Specifically, the DT always performs the first inference on each new input datum, looking for \textit{easy} activity classes. For example, inputs corresponding to the ``laying'' activity can be intuitively recognized looking just at the orientation and amount of movement detected by the accelerometer. To cover more complex activities, the DT is trained to output a generic ``fallback'' class, which can be interpreted as: ``the input does not belong to any of the easy classes''.
When the DT output is different from ``fallback'', the inference terminates immediately, using the predicted class as final result. On the contrary, when the fallback class is predicted, the 1D CNN is activated and applied to the same input, in order to distinguish among more complex activities (e.g., stairs climbing, running, cycling, etc).

The average energy consumption of this hierarchical model can be computed with the following equation:
\begin{equation}
    E=E_{DT} + P_{fallback} * E_{CNN}
\end{equation}\label{eq:hierarchy}
where $E_x$ denotes the inference energy of $x$, and $P_{fallback}$ denotes the probability of an input being classified as ''fallback'' by the DT, thus enabling the CNN. Since the DT
is significantly less computationally expensive than the CNN ($E_{DT} << E_{CNN}$), large energy savings are obtained for easy inputs, which only require the execution of the first ``stage''.
Moreover, the latter are often the majority in real-life scenarios (we sleep for more hours than we jump or climb stairs), meaning that the \textit{total} savings will also be considerable.

Additionally, with respect to a standard, single-stage classifier, our second-stage CNN operates on a reduced set of classes (only the complex ones), which reduces both the number of inference operations and the number of parameters of last fully-connected layer. The former further cuts the energy cost of the system, while the latter helps compensating the memory overhead introduced by the presence of the DT.

\subsection{Training and Deployment Flow}\label{sec:flow}

Fig.~\ref{fig:workflow} shows the training and deployment flow that we propose for the two-stage adaptive system just described, which is composed of five main steps.

\subsubsection*{1. DT-based Task Decomposition}
In the first phase, we identify the easy and hard classes that will be later assigned to the DT and CNN respectively. To this end, we train and optimize the hyper-parameters of a DT on the \textit{full} task (i.e., on all $M$ activity classes). Namely, we select the tree depth that yields the best classification results on the full training set, by means of an exhaustive grid search, made possible by the limited size of HAR datasets. In our experiments, we explore depth values in [2,10]. Then, we select as easy classes the $M_{easy} < M$ static activities for which the complete DT achieves the largest F1 score. In general, the number of easy classes is a design parameter that should be explored, but in this preliminary work, we fix it to $M_{easy} = 2$. All other classes are considered hard.

\subsubsection*{2. Sub-tasks Training Sets Generation}
Next, we generate the training sets for the DT and CNN. The former is obtained keeping samples relative to all easy classes unchanged, and modifying hard samples so that they all have the same label, corresponding to the fallback class. The CNN training set, instead, is simply obtained removing all easy classes samples, since those are not considered in the second inference stage.

\subsubsection*{3. Final DT Selection and Training}
We then perform a second set of trainings and depth optimizations on the DT, this time on the modified training set containing only easy-classes and fallback examples. We look once again for the model that maximizes the classification score, using grid search.

\subsubsection*{4. CNN Pareto Exploration}
For what concerns the CNN, we perform a similar search, but in this case rather than only selecting the most accurate model, we extract the entire Pareto frontier in the energy versus classification score space. This difference comes from the fact that the second stage of the hierarchical model is by far the most computationally expensive. Therefore, building the hierarchical system starting from different Pareto-optimal CNNs is the easiest way to obtain different accuracy versus energy consumption trade-offs, as shown in Sec.~\ref{sec:results}.
The CNN architecture exploration is performed starting from a common template, based on a classic network structure~\cite{lenet}, composed of three convolutional blocks (each including a convolution, batch normalization and max-pooling) and a final fully-connected layer. We explore variants of this template obtained changing the number of channels and the kernel size in each convolutional layer. Specifically, we consider all possible combinations of power-of-2 channels in $[2,...,128]$ and kernel sizes 7 or 15. The max-pooling stride and width are left fixed at 2.

\subsubsection*{5. MCU Deployment}
Lastly, we deploy the selected DT and one of the Pareto-optimal CNNs onto the target MCU, combining them with the scheme of Fig~\ref{fig:architettura}, using optimized libraries (described below) to implement both classifiers.

\section{Experimental Results}\label{sec:results}

We target a popular public HAR dataset named UCI HAPT~\cite{hapt}, featuring 958500 samples collected from the accelerometer and gyroscope of an Android phone. The data has been collected with a sampling frequency of 50 Hz from 30 subject performing 12 different activities. 
We keep the same train/test split proposed by the authors and split the data in non overlapping windows of 5 s, obtaining inputs of size 250x6.
For this dataset, the 2 classes identified as easy by our procedure, hence those classified by the DT, are \textit{sitting} and \textit{laying}, which result in the highest F1-score (above 85\%) on the training set.
Together, these two activities account for $\approx$ 30\% of both training and testing samples.

As deployment target, we select the single-core, ultra-low-power, RISC-V MCU Quentin~\cite{quentin}, with characteristics close to those of common embedded HAR devices. The MCU has 520 kB of memory, and our results are obtained with the core running at 205.1 MHz, with a supply voltage of 0.54 V. In these conditions, the system consumes 3.8 mW of active power.

We train our DTs and CNNs in Python, before converting the trained models to optimized C code for the target MCU. For this conversion, we use the library described in~\cite{Burrello2021b} for the 1D CNN, while the DT and RF implementations are described in~\cite{daghero2021ultracompact}.
Both libraries are specifically tailored for the target hardware, leveraging the available SIMD and DSP-oriented instruction set extensions.

\subsection{Energy and memory comparison}

\begin{figure}[t]
\centering
\includegraphics[width=\linewidth]{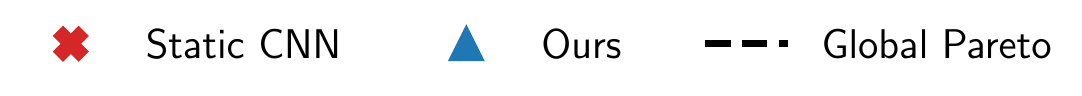}

\begin{subfigure}{\linewidth}%
\includegraphics[width=0.47\linewidth]{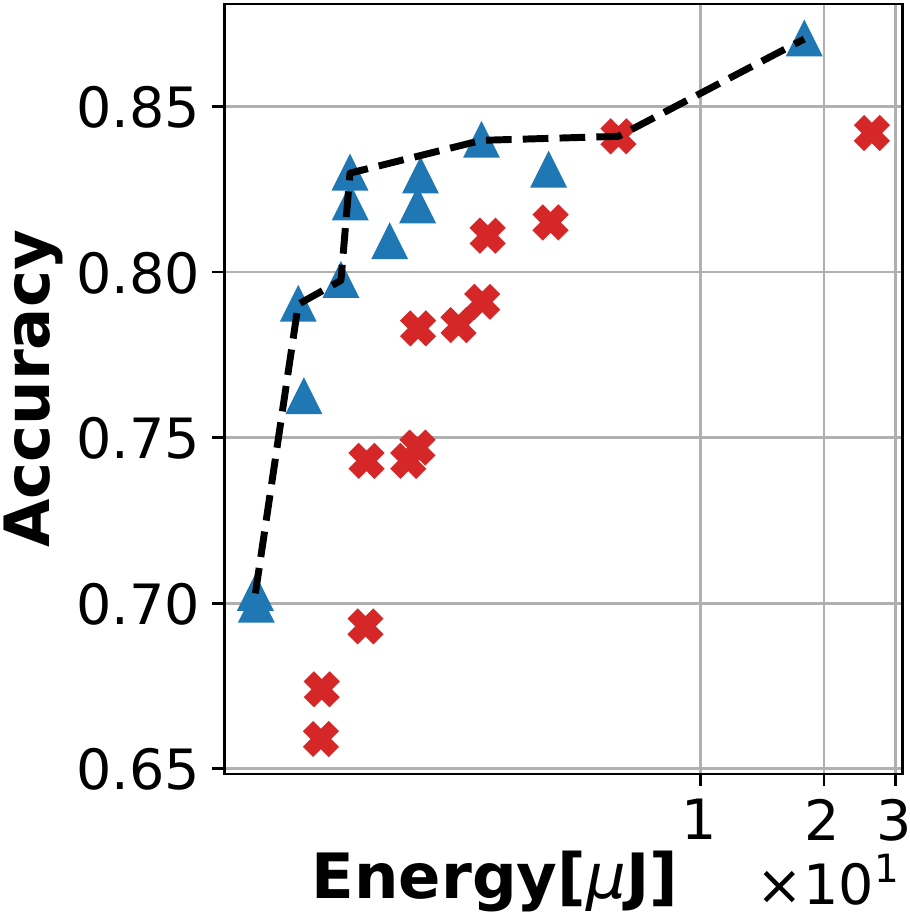}%
\includegraphics[width=0.47\linewidth]{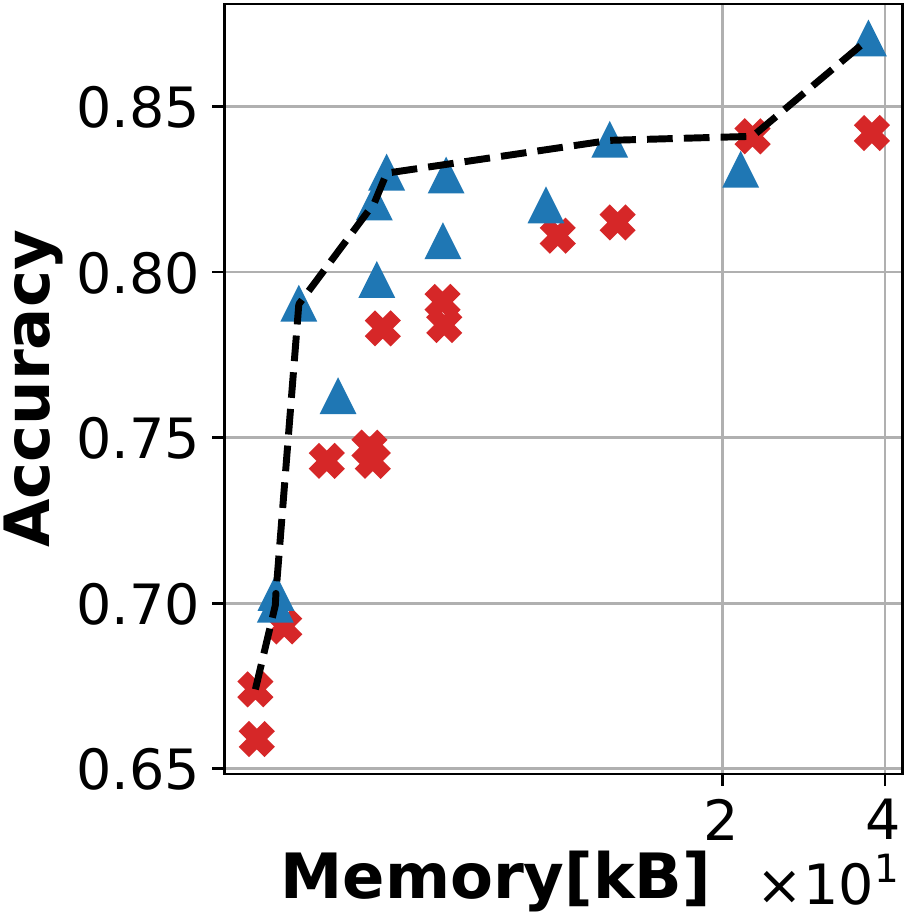}%
\end{subfigure}%
\caption{
Accuracy versus average energy consumption and total memory occupation results, for the proposed two-stage method and for a static CNN-based approach.
}
\label{fig:adaptive_vs_static}
\end{figure}

Fig.~\ref{fig:adaptive_vs_static} shows the results  obtained by our two-stage architecture in terms of classification accuracy versus inference energy consumption and memory occupation on the target MCU.
Specifically, each blue triangle represents a hierarchical system built combining the DT determined in Step 3 of Fig.~\ref{fig:workflow} (always the same) and one of the different Pareto-optimal CNN architectures found in Step 4 of Sec.~\ref{sec:flow}. For comparison, the two graphs also report the results obtained with different \textit{static} CNNs derived from the same search space, and trained directly on \textit{all} activity classes (red crosses). The dashed black line identifies the global Pareto front. Energy results refer to the average consumption for a single inference, over the entire test set.

On the energy side, Fig.~\ref{fig:adaptive_vs_static} shows that our approach occupies most of the Pareto frontier, with the exception of a single architecture around 85\% accuracy. Namely, we are able to reduce the average energy per inference by up to 67.7\% at iso-accuracy, or to improve the accuracy by up to 12.34\% for the same average energy. Further, our most accurate adaptive system outperforms the best static CNN (+3\% accuracy) with lower energy consumption.
This result is obtained thanks to the fact that the DT achieves a classification accuracy of 92\% on the ``easy activities'' sub-task, while having negligible inference energy costs (341x less than the smallest CNN).

Memory results are similar, showing that the presence of an additional DT in our two-stage system yields negligible overheads. Specifically, the DT requires a total of 0.2kB of memory, yielding a maximum overhead of around 9\% w.r.t to the smallest static CNN ($ \leq 3$ kB). 
At the opposite side of the Pareto front, thanks to the reduced dimension of the last CNN fully-connected layer (given by the reduced number of activities classified), the total memory of the system is instead reduced by up to 64\% for the same accuracy.

\begin{figure}[t]
\centering
\includegraphics[width=0.7\linewidth]{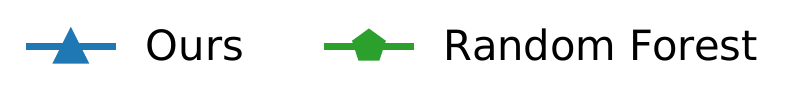}

\begin{subfigure}{\linewidth}%
\includegraphics[width=0.47\linewidth]{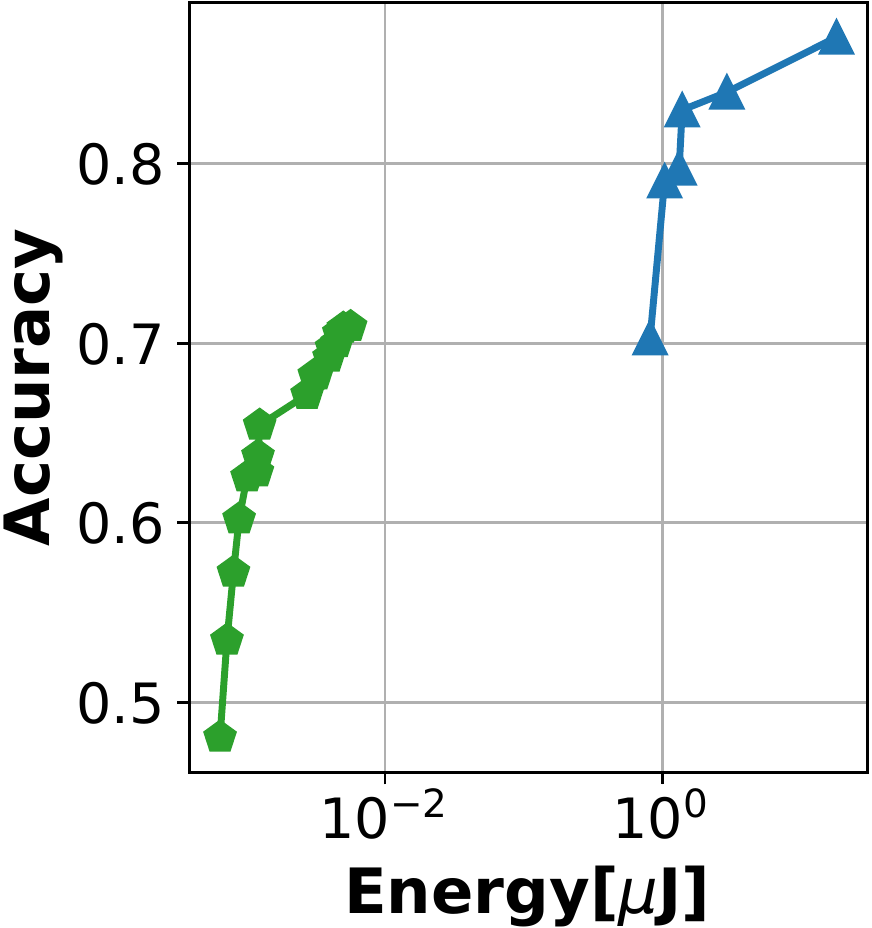}%
\includegraphics[width=0.47\linewidth]{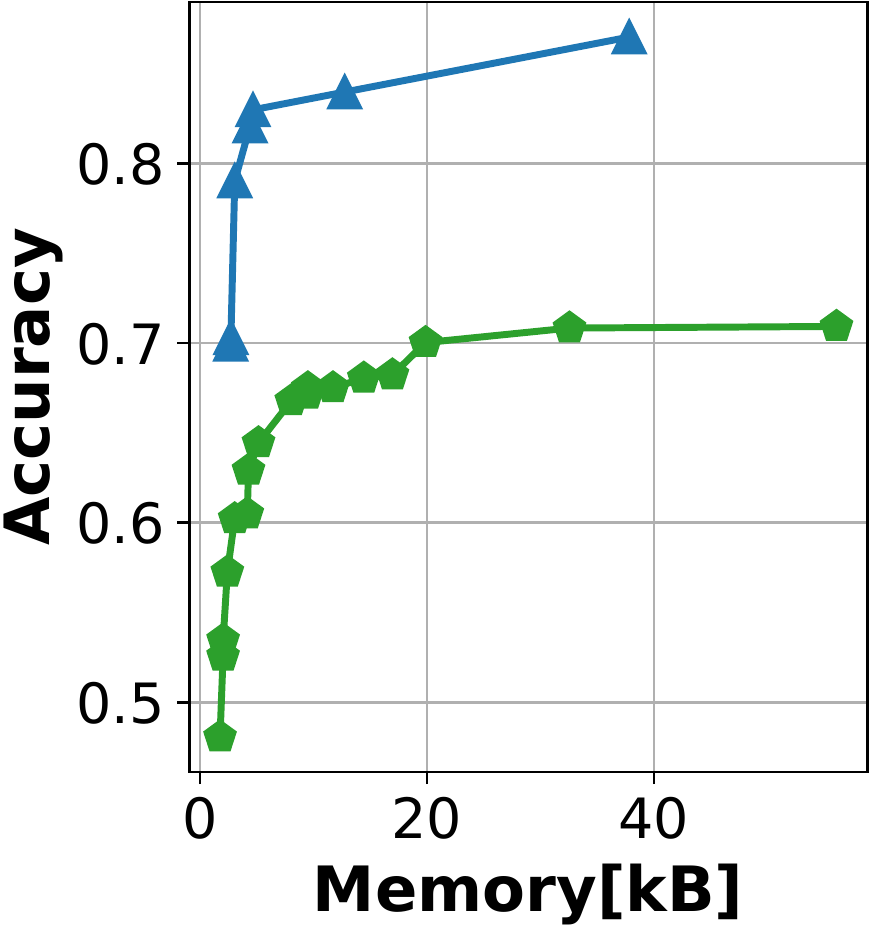}%
\end{subfigure}%
\caption{
Accuracy versus average energy consumption and total memory occupation results for the proposed two-stage method and for a Random Forest classifier.
}
\label{fig:rf_comparison}
\end{figure}

Figure~\ref{fig:rf_comparison} compares our method with a fully tree-based solution, consisting either of a single DT or of a Random Forest (RF) ensemble~\cite{daghero2021ultracompact}.
Precisely, the green curve shows the Pareto front obtained training all RFs with a number of trees from 1 (single DT) to 15, and depths in the $[2,20]$ interval. The blue curve is the Pareto front of our adaptive method.
As explained in Sec.~\ref{sec:intro}, tree-based models are expected to have lower energy consumption than DL ones, due to their lightweight computations, as confirmed by the graphs.
However, also as expected, these models are unable to achieve comparable accuracy to a DL-based method within the memory constraints of a typical MCU. Indeed, even the largest RFs, which occupy more memory than our adaptive system, cannot achieve more than 71\% accuracy.

\subsection{Detailed Deployment Results}

Table~\ref{tab:deployment} reports the detailed deployment results, in terms of memory occupation, energy consumption, and inference latency, of the most accurate version of our adaptive system, and of the most accurate static CNN (\textit{Base} rows). The results show that, while obtaining a +3\% accuracy increase, the adaptive system can reduce the average energy and latency per inference by 31.6\% even at maximum accuracy.
Additionally, in order to show the potential advantages of such a system in a real-world scenario, we repeat the experiment after synthetically over-sampling \textit{laying} and \textit{sitting} inputs (i.e., the easy classes) in the test set by 10x and 20x. In those settings, the accuracy improvement reaches +4\%, and the energy/latency reductions increase to 76\% and 83\% respectively.

\begin{table}[t]
\begin{tabular}{llllll}
\textbf{} & \textbf{Mode} & \textbf{Acc.} & \textbf{Memory[kB]} & \textbf{Energy[$\mu$j]} & \textbf{Latency[ms]} \\ \hline \hline
\multirow{2}{*}{Base} & Ours   & 0.87 & 37.2 & 17.9 & 4.7 \\ 
                      & Static & 0.84 & 37.8 & 26.2 & 6.9 \\ \hline
\multirow{2}{*}{10x}  & Ours   & 0.89 & 37.2 & 6.2  & 1.6 \\
                      & Static & 0.85 & 37.8 & 26.2 & 6.9 \\\hline
\multirow{2}{*}{20x}  & Ours   & 0.90 & 37.2 & 4.4  & 1.2 \\
                      & Static & 0.86 & 37.8 & 26.2 & 6.9 \\ \hline
\end{tabular}
\caption{Detailed deployment results on the original and synthetically augmented test data.}\label{tab:deployment}
\end{table}

\section{Conclusions}
We have presented a novel, two-stage HAR system for MCUs, featuring a small DT tree cascaded with a 1-dimensional CNN. Our approach leverages the different input complexity to save energy at inference time, switching on the CNN only when needed. Compared to a static CNN, we reduce the average energy per inference by up to 67.7\% at iso-accuracy on a state-of-the-art HAR dataset. With respect to a pure tree-based model, our approach occupies a different portion of the design space, yielding much higher accuracy for the same memory occupation.

\bibliographystyle{IEEEtran}
\bibliography{library}

\end{document}